\documentclass{article}
\usepackage{spconf,amsmath,graphicx}
\usepackage{cite}
\usepackage{amsmath,amssymb,amsfonts,tikz}
\usepackage{graphicx}
\usepackage{textcomp}
\usepackage{xcolor}
\usepackage{main}
\usepackage{multirow}
\usepackage{bm}
\usepackage{bbold}
\usepackage{cite}
\usepackage{subfig}
\usepackage{url}
\usepackage{nicefrac}

\usetikzlibrary{matrix}

\graphicspath{{./Figures/}}

\title{Using Received Power in Microphone Arrays to \\Estimate Direction of Arrival}

\name{Gustav Zetterqvist \qquad Fredrik Gustafsson \qquad Gustaf Hendeby \thanks{G. Zetterqvist has received funding from ELLIIT. 
This work was partially funded by the Wallenberg AI, Autonomous Systems and Software Program (WASP) funded by the Knut and Alice Wallenberg Foundation.
The authors would like to thank Jonas Nordlöf at the Swedish Defence Research Agency (FOI) for the help and assistance during the data collection.}
}

\address{Division of Automatic Control, Linköping University, SE-581 83 Linköping, Sweden}

\begin{document}

\maketitle

\begin{abstract}
Conventional \emph{direction of arrival} (DOA) estimators are based on array processing using either time differences or beamforming. 
The proposed approach is based on the received power at each microphone, which enables simple hardware, low sampling frequency and small arrays. 
The problem is recast into a linear regression framework where the least squares method applies, and the main drawback is that different sound sources are not readily separable. 

Our proposed approach is based on a training phase where the directional sensitivity of each microphone element is estimated. 
This model is then used as a fingerprint of the observed power vector in a real-time estimator. 
The learned power vector is here modeled by a Fourier series expansion, which enables Cramér-Rao lower bound computations.
We demonstrate the performance using a circular array with eight microphones with promising results.
\end{abstract}

\begin{keywords}
DOA Estimation, Directional Sensitivity, Microphone Array, CRLB, YALMIP
\end{keywords}

\section{Introduction}
\emph{Direction of arrival} (DOA) estimation is a well studied topic during the last decades \cite{Knapp:DoaStudy,Krim:DoaStudy,Stoica:DoaStudy}, 
see \cite{Rascon:RecentDoaStudy} for a more recent study. 
Conventional DOA estimation is based on estimating the time delay for the received signal waveform. 
This principle is the dominating one for all kinds of arrays, whether it is acoustic, microwaves or sonar waves. 
The most general approaches with no constraints on the array geometry are based on estimating the time delays to get a set of \emph{time-difference of arrival} (TDOA) measurements and then solving a nonlinear least squares problem, 
\emph{ e.g.} \emph{multiple signal classification} (MUSIC) \cite{Schmidt:MUSIC} and \emph{minimum variance distortionless response} (MVDR) \cite{Capon:MVDR)}. 
This principle can be applied to essentially any array geometry, any signal wave form and works both in the near-field and far-field. 
Conversely, assuming a far-field planar wave with a narrowband signal opens up for frequency domain approaches. 
Further, uniform linear or circular arrays enable efficient algorithms.

DOA estimation is conveniently included in sensor fusion applications. 
A network of arrays can enable object localization by triangulation. 
More generally, a DOA estimator can be seen as an angle-only measurement in an object tracking system.

Our approach is based on a training phase to learn the directional sensitivity of each sensor element. 
This angular response can conveniently be approximated by a low-order parametric model, where we advocate a \emph{Fourier series} (FS) expansion. 
In the DOA estimation phase, the measured power profile is then matched to the model using the least squares method. 
The parametric model enables analytical expressions for the \emph{Cram\'er-Rao lower bound} (CRLB). 
This can be a convenient tool for evaluating a particular array design.

DOA estimation only using the directional sensitivity of the sensor elements has a number of advantages \cite{Buck:Directivity,Nakadai:Directivity}.
Firstly, the sampling frequency can be considerably relaxed, thus decreasing the computational burden of the array.
Secondly, there exists low-cost microphones that only measure the power, not the samples, enabling affordable arrays and large scale deployments. 
In fact, most microphone signal processing can be performed in the analogue domain using simple electronics.
Also, the theory simplifies a lot, since standard \emph{least squares} (LS) can be applied, which is suitable for energy efficient microprocessor implementations in the edge.
Finally, the array can be made arbitrarily small, since the distance between the sensors does not matter and nonuniform arrays do not add any extra complications.

The drawbacks are mainly that a training phase is needed and that there are certain requirements on the array design. 
For instance, an array with all sensors aligned in the same direction is pointless, since there is no directional sensitivity in that case.
Another drawback is that we can not separate different signal sources.

Even though the power approach is general to all signals (acoustic, sonar, radio, seismic), we are mainly focusing on acoustic applications using microphone arrays. 
Microphones have a natural directional sensitivity, which can be further improved by the design. 
We present promising results based on a mockup array. 

\section{Signal Model}
Denote the measured signal from microphone~$n$ at discrete time~$l$ with the scalar
\begin{equation}\label{eq:signal}
    y_n(l) =  s_n(l) + w_n(l) \mathrm{,}
\end{equation}
where $s_n(l)$ is the received signal and $w_n(l)$ is measurement noise uncorrelated to $s_n(l)$, and is assumed to be normally distributed white noise, $w_n(l) \sim \mathcal{N}(0,\sigma_n^2)$.
The power of the measured signal from microphone~$n$ can then be calculated as
\begin{equation}\label{eq:power}
    P_n = \frac{1}{L} \sum_{l=1}^L y_n(l)^2 \mathrm{,}
\end{equation}
where $L$ is the number of samples in the signal of interest.
By inserting \eqref{eq:signal} into \eqref{eq:power}, the signal power can thus be expressed in three terms
\begin{equation}\label{eq:power_terms}
    P_n = \underbrace{\frac{1}{L} \sum_{l=1}^L s_n^2(l)}_{ P^s_n} + 
    \underbrace{\frac{1}{L} \sum_{l=1}^L 2 s_n(l) w_n (l)}_{P^{sw}_n} + 
    \underbrace{\frac{1}{L} \sum_{l=1}^L w_n^2(l)}_{e_n}  \mathrm{,}
\end{equation}
where $P^s_n$ is the power of the received signal, $P^{sw}_n$ is a cross-term from the signal and the noise, and $e_n$ is the power of the measurement noise.
Here, the number of samples $L$ is assumed large, hence $P_n^{sw}\rightarrow 0$. 
Further, since $w_n(t)$ is normally distributed, $e_n$ will be chi-squared distributed with $L$ degrees of freedom, $\frac{L}{\sigma_n^2} e_n \sim \chi_{L}^2$.
However, since the degrees of freedom is assumed to be large the chi-squared distribution is approximated with a normal distribution 
\begin{equation}\label{eq:noise_dist}
    \frac{L}{\sigma_n^2}e_n \sim \chi_{L}^2 \stackrel{\mathrm{Approx}}{\rightarrow} e_n \sim \mathcal{N}(\sigma_n^2,\frac{2\sigma_n^4}{L}) \mathrm{.}
\end{equation}

The goal is to estimate the DOA, denoted as $\psi$. 
In a tracking framework, $\psi$ can be interpreted as the heading to the object, and the DOA estimator is an angle-only sensor. 

We assume that each microphone has a directional sensitivity in the power attenuation by design or construction of the array. 
The absolute level of the received power, assumed to be the same at all sensors, is denoted $\alpha$ and is considered to be a nuisance parameter from a DOA estimation perspective. 
Thus, the power $P_n$ measured by microphone~$n$ can be expressed as
\begin{align}
P_n(\psi) = \alpha g_n h(\psi,\bm{\theta}_n) + e_n \mathrm{,}
\end{align}
where $g_n$ is the microphone gain, 
$h(\psi,\bm{\theta}_n)$ is the directional sensitivity of the microphone dependent on the DOA angle $\psi$ and parametrized by the parameters $\bm{\theta}_n$, 
and $e_n$ is the observation error as described in \eqref{eq:noise_dist}.

\section{Method}
The approach we propose is based on two steps, training and estimation.

\subsection{Training}
In the training phase, the array is exposed to wideband noise from different directions $\psi$ distributed around the array, and $y_n$ is observed in a controlled environment. 
Then, the power of the measured signal, $P_n(\psi)$, is calculated according to \eqref{eq:power}.

The parameters of the directional sensitivity~$\bm{\theta}_n$ as well as the microphone gain~$g_n$ are thereafter estimated using YALMIP with the FMINCON solver \cite{Lofberg2004}.
The optimization problem is 
\begin{equation}\label{eq:optimization_problem}
    \begin{aligned}
    & \underset{\bm{x}}{\text{minimize}}
    & & V(\bm{x}) \\
    & \text{subject to}
    && \alpha > 0 \\
    &&& g_n > 0 & \hspace{-0 cm} \forall \, n = 1,2, \dots , N \\
    &&& h(\psi_n,\bm{\theta}_n) = 1 & \hspace{-0 cm} \forall \, n = 1,2, \dots , N \\
    &&& \textstyle \sum_{n=1}^N g_n^2 = 1
    \end{aligned}
\end{equation}
where $\psi_n$ is the angle when microphone~$n$ is facing the sound source, $N$ is the number of microphones and $V(\bm{x})$ is the loss function
\begin{equation}
    V(\bm{x}) = \sum_{n=1}^N \frac{L}{2\sigma_n^4}\sum_{k=1}^K \Big( P_n(\psi_k) - \left(\alpha g_n h(\psi_k, \bm{\theta}_n) + \sigma_n^2\right) \Big)^2 \mathrm{,}
\end{equation}
where $\bm{x}$ contains the optimization variables $\alpha$, $\{g_1, \dots, g_N\}$ and $\{\bm{\theta}_1, \dots, \bm{\theta}_N\}$, and $K$ is the number of observed directions.
The term $\sigma_n^2$ appears since the noise does not have zero mean, and is estimated with measurements of background noise.

\subsection{Fourier Series Model}
To model the directional sensitivity $h(\psi,\bm{\theta}_n)$ for microphone~$n$, we adopt a FS model 
\begin{align}
h(\psi,\bm{\theta}_n)  &= \theta^n_{0} + \sum_{d=1}^D \theta^n_{d,c} \cos(d\psi)+\theta^n_{d,s} \sin(d\psi) \nonumber \\
&= \bm{\Phi}(\psi) \bm{\theta}_n \mathrm{,}
\end{align}
where $D$ is the order of the FS.
The order of the FS expansion is then determined using the \emph{Bayesian information criterion} (BIC) \cite{Ljung:SI}.

\subsection{Estimation}
From the training, the model for microphone~$n$ can be expressed as
\begin{align}
    \hat{P}_n(\psi) = \alpha \hat{g}_n \bm{\Phi}(\psi) \bm{\hat{\theta}}_n + \sigma^2_n \mathrm{,}
\end{align}
where $\hat{g}_n$ is the estimated microphone gain and $\bm{\hat{\theta}}_n$ is the estimated parameters from the training phase.

The DOA estimator can then be computed using LS as
\begin{align}\label{eq:LS_estimation}
    \hat{\psi}  &= \arg\min_{\psi} \sum_{n=1}^N \left( \frac{P_n - \sigma_n^2}{\| \bm{P} \|} - 
    \frac{\hat{P}_n(\psi) - \sigma_n^2}{\| \bm{\hat{P}}(\psi) \|}\right)^2 \mathrm{, }
    \\ \nonumber
    \| \bm{P} \| &= \sqrt{\sum_{n=1}^N \left(P_n - \sigma_n^2\right)^2} \mathrm{, } \;
    \| \bm{\hat{P}}(\psi) \| = \sqrt{\sum_{n=1}^N \left(\hat{P}_n(\psi) - \sigma_n^2\right)^2}
\end{align}
where the normalization is performed in order to only estimate the direction and not the strength of the signal.
Thus, $\alpha$ is seen as a nuisance parameter in the DOA estimate.

\section{Cram\'er-Rao lower bound}
To calculate the CRLB a \emph{probability density function} (PDF) of the measurement is required. 
Since the noise $\bm{e}$ is assumed normally distributed as $\bm{e} \sim \mathcal{N}(\sigma^2 \mathbb{1},\lambda \bm{I})$, 
the PDF of the measurement is denoted as 
\begin{equation}
    p(\bm{P}|\psi,\bm{\theta},\alpha,\lambda) = \mathcal{N}( \alpha\bm{G}\bm{h}(\psi,\bm{\theta})+ \sigma^2 \mathbb{1}, \lambda \bm{I}) \rm{,}
\end{equation}
where $\bm{P}$ is a vector with the power measurement of all microphones, $\bm{G}$ is the gain matrix with each microphone's gain~$g_n$ on the diagonal, 
$\bm{h}(\psi,\bm{\theta})$ is the directional sensitivity of all microphones, i.e. each $h(\psi,\bm{\theta}_n)$ stacked in a vector, 
and $\lambda=\nicefrac{2\sigma^4}{L}$ is the noise variance assumed to be the same for all microphones.

This yields the \emph{Fisher Information Matrix} (FIM)
\begin{align}
    \mathcal{I}(\psi,\alpha) &= - \mathrm{E} \left[ 
        \begin{pmatrix}
            \frac{\partial^2 }{\partial \psi^2} & \frac{\partial^2 }{\partial \psi \partial \alpha} \\
            \frac{\partial^2 }{\partial\alpha \partial\psi} & \frac{\partial^2 }{\partial \alpha^2} 
        \end{pmatrix}
        \log p(\bm{P}|\psi,\bm{\theta},\alpha,\lambda)\right] \nonumber \\ 
    &= \frac{1}{\lambda} 
    \begin{bmatrix}
       \alpha^2 \bm{h}'^{\;T}_\psi \bm{G}^T \bm{G} \bm{h}'_\psi  &\alpha \bm{h}'^{\;T}_\psi \bm{G}^T\bm{G} \bm{h} \\\
       \alpha \bm{h}'^{\;T}_\psi \bm{G}^T \bm{G} \bm{h} & \bm{h}^T \bm{G}^T \bm{G}\bm{h}
    \end{bmatrix}\rm{,}
\end{align}
where $\bm{h}$ denotes $\bm{h}(\psi,\bm{\theta})$ and $\bm{h}'_\psi = \frac{\partial \bm{h}(\psi,\bm{\theta})}{\partial \psi}$.

Assuming that the estimate in \eqref{eq:LS_estimation} is unbiased and that the PDF $p(\bm{P}|\psi)$ satisfies the regularity conditions,
the CRLB can be used to evaluate the performance of the estimation method \cite{kay:estimation}.
Thus, the CRLB of $\psi$ is computed as
\begin{multline}
    \mathrm{var}(\hat{\psi}) \geq \mathrm{CRLB}(\psi) = \left[\mathcal{I}(\psi, \alpha)^{-1}\right]_{1,1}  \\
    = \frac{\lambda}{\alpha^2}\frac{\|\bm{G}\bm{h}\|^2}{\left(\|\bm{G}\bm{h}\|^2 \: \|\bm{G}\bm{h}'_\psi\|^2 - \langle \bm{G}\bm{h}, \bm{G}\bm{h}'_\psi \rangle^2 \right)}  \rm{,}
\end{multline}
where the denominator is strictly positive as long as $\bm{h}$ and $\bm{h}'_\psi$ are not parallel according to Cauchy–Schwarz inequality. 
The term $\alpha^2/\lambda$ can be interpreted as the \emph{signal-to-noise ratio} (SNR) of the signal. 

\section{Experimental Array Design}
To evaluate the method, a hardware prototype has been designed and used to collect data in an anechoic chamber.
\subsection{Hardware Prototype}
\begin{figure}[tb]
    \centering
    \subfloat[][The sensor array in the anechoic chamber. \label{fig:array_overview}]{\includegraphics[width=0.4\columnwidth]{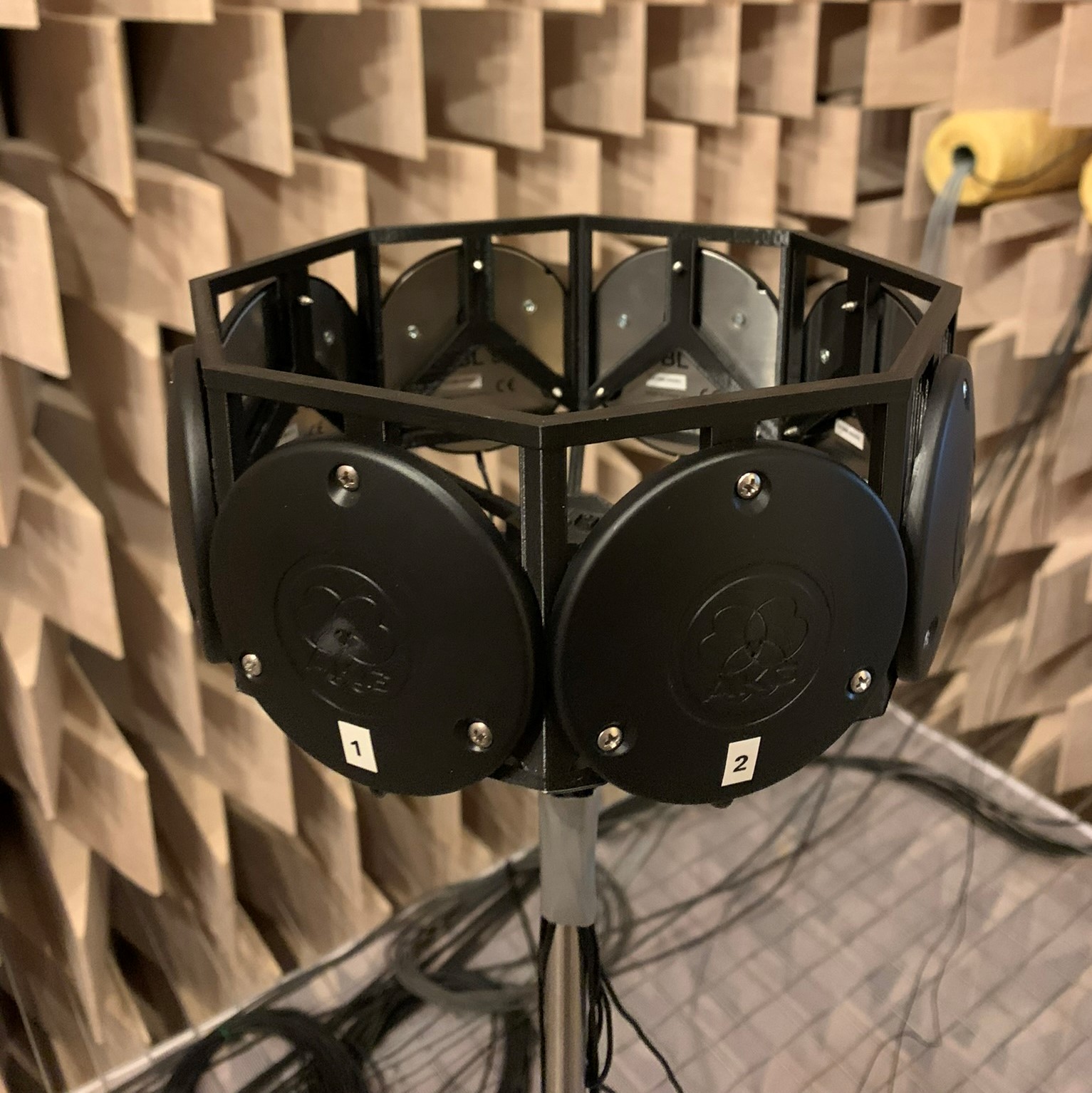}}
    \hspace*{0.02\columnwidth}
    \subfloat[][Top view of the dimension and microphone placement of the sensor array. \label{fig:array_dim}]{\includegraphics[width=0.4\columnwidth]{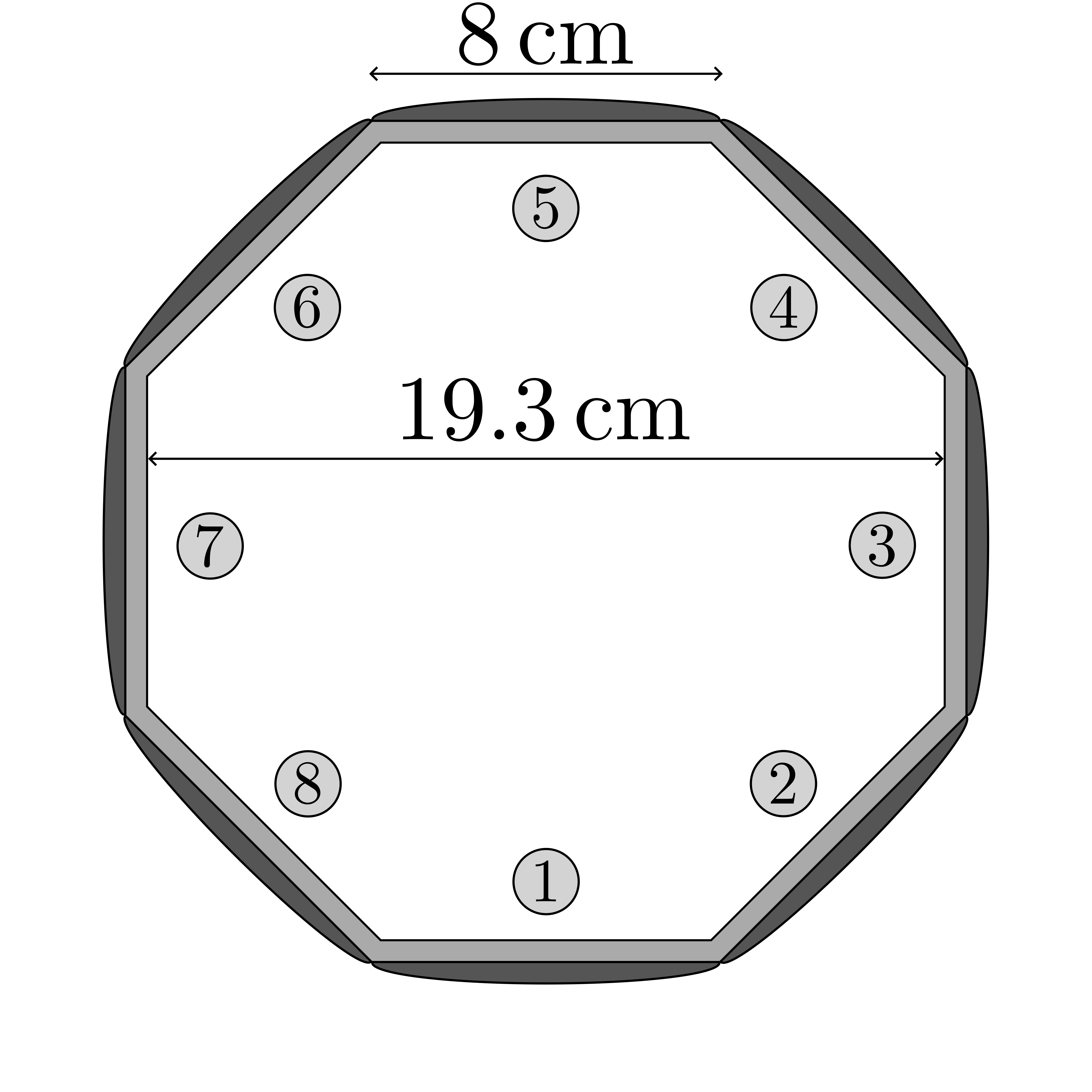}}\\
    \caption{An overview of the sensor array used in this paper.}
    \label{fig:array}
\end{figure}

The sensor array used in this paper is a \emph{uniform circular array} (UCA) in the form of an octagon with eight microphones.
The microphones used are omnidirectional CBL99 from AKG \cite{CBL99} connected to a Behringer UMC1820 preamplifier \cite{UMC1820}.
The preamplifier is connected to a computer which allows for a sampling frequency of up to $96$\,kHz. 
The array and its dimensions are shown in \Figref{fig:array}.

\subsection{Measurements of Directional Sensitivity}
Data were collected in an anechoic chamber at the Swedish Defence Research Agency (FOI) in Linköping, ISO rated from $100$\,Hz to $10$\,kHz. 
The sensor array was put on a turntable and a Genelec~1029A speaker \cite{1029A} was placed 5 meters away.
In the setup microphone~1 was facing the speaker at $\psi=0$\textdegree{}, microphone~2 at $\psi=-45$\textdegree{} and so on.
Thereafter, data were recorded with a sampling frequency of $48$\,kHz using 5 different types of signals.
(1) \textit{Wideband noise} with a bandwidth of $100$\,Hz--$10$\,kHz that lasted for 10 seconds.
(2) \textit{A hovering drone} sound during 5 seconds.
(3) \textit{Attenuated wideband noise}, i.e. the wideband noise signal with half the amplitude.
(4) \textit{Amplified wideband noise}, i.e. the wideband noise signal with doubled amplitude.
(5) \textit{Background noise} collected during 26.41 seconds.

\section{Estimation results}
In this section the results from the method is presented using the collected data.
\subsection{Training}\label{sec:results_training}
For the training, the wideband noise signal is measured from 24 angles uniformly distributed around the array.
Then, the signal power of the measured signals is calculated according to \eqref{eq:power}. 
Thereafter, the noise variance $\sigma_n^2$ is calculated from measurements of the background noise.
Finally, the optimization problem~\eqref{eq:optimization_problem} is solved to estimate the model parameters.

\subsection{Fourier Series Fit}
\begin{figure}[tb]
    \centering
    \includegraphics[width=0.55\columnwidth,trim={0cm 3.075cm 0cm 0cm},clip]{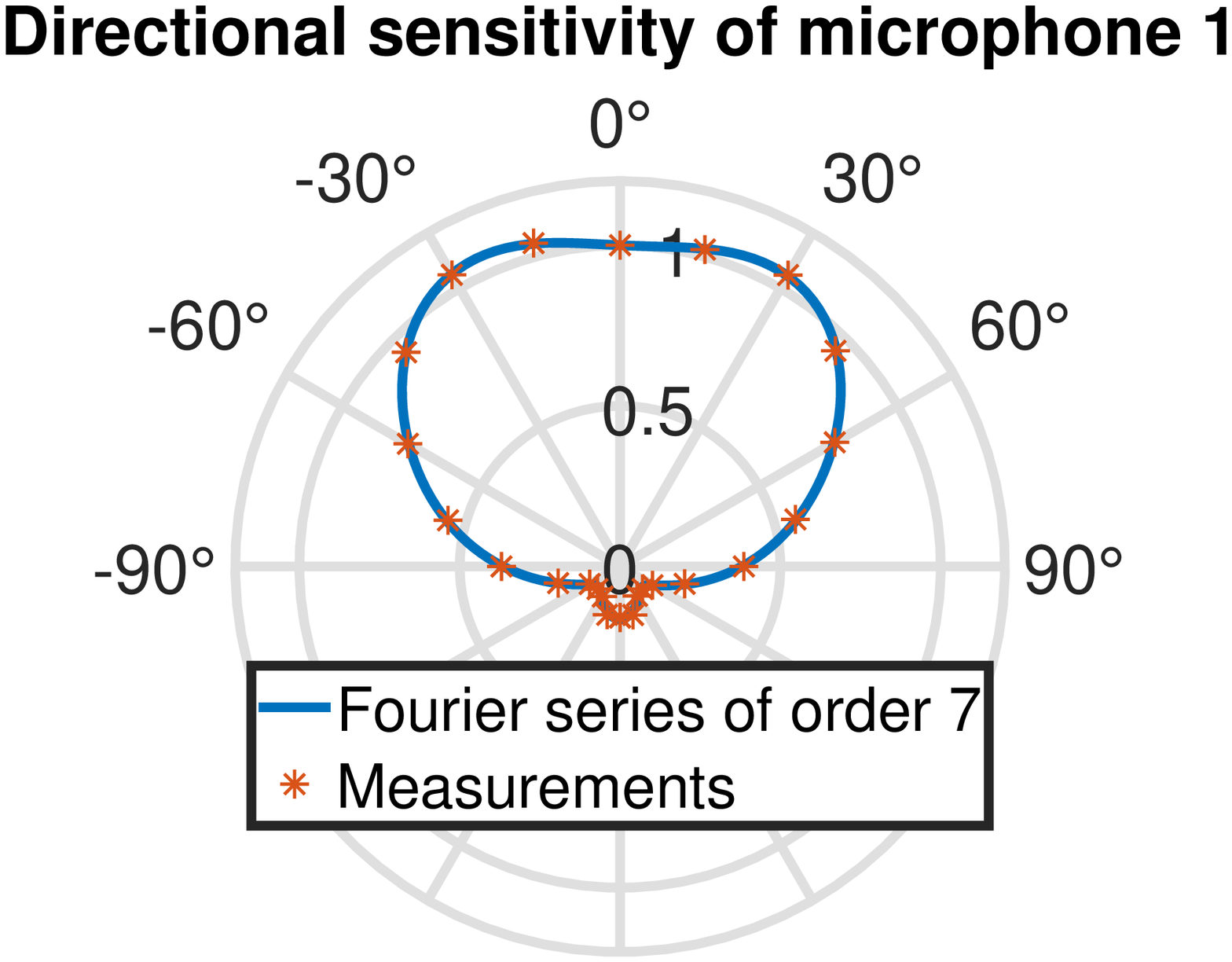}
    \caption{The measured power of the wideband noise signal normalized with the power at $\psi=0$\textdegree{} and the FS of order~7 as a function of the DOA for microphone~1.}
    \label{fig:norm_power_wn_fourier_2}
\end{figure}

In the optimization problem different orders of the FS are examined to find the best one.
The best model order is determined using the BIC, which indicates that a model order of~7 is a good trade-off. 
The measured power of the wideband noise signal normalized with the power at $\psi=0$\textdegree{} as well as the corresponding FS of order~7 is illustrated in \Figref{fig:norm_power_wn_fourier_2}.

\subsection{Cram\'er-Rao lower bound}
\begin{figure}[tb]
    \centering
    \includegraphics[width=0.95\columnwidth]{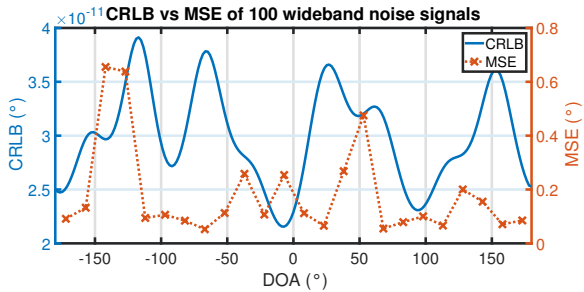}
    \caption{The calculated CRLB for the wideband noise signal as well as the MSE of the wideband noise signal divided into 100 subsignals using a model order of~7.}
    \label{fig:CRLB_vs_mean_error_wn}
\end{figure}

By using the calculated noise variance and the resulting $\alpha$ from the training step, 
a SNR of $130$\,dB is achieved, and the resulting CRLB is shown in \Figref{fig:CRLB_vs_mean_error_wn}.

\subsection{DOA Estimation}
\begin{figure}[tb]
    \centering
    \includegraphics[width=0.9\columnwidth,trim={2cm 0cm 2cm 0.5cm},clip]{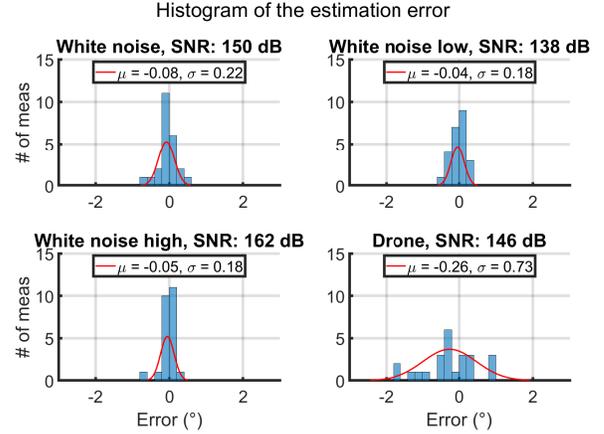}
    \caption{Histogram of the estimation error for the validation data with a FS expansion of order~7. The red line is a normal distribution fit to the estimation error.}
    \label{fig:hist_fourier}
\end{figure}

For the DOA estimation, all signals are used except the background noise.
The validation data is collected at 24 angles uniformly distributed around the microphone array, but at different angles than the training data.
Then, the power of the signals is calculated and compared to the estimated FS model using the LS from \eqref{eq:LS_estimation}.
A histogram of the estimation error with a FS of order~7 is presented in \Figref{fig:hist_fourier}.

Here, it is seen that all signals are estimated with good precision. 
It is also noted that the SNR does not seem to affect the result in any significant way.
This suggests that the main source of estimation error is not the measurement noise, but it could be due to a model error. 
Also, the drone performs worse than the wideband noise signals.
The reason for this is unclear since the signal power is quite high, but it could be caused by the difference in the frequency content of the signal.

By looking at the SNR values, they appear to be extremely high.
The reason for this is probably that the variance of the observation error is higher than estimated or that the variance of the cross term in~\eqref{eq:power_terms} is not negligible.

Further, by looking at the \emph{mean squared estimation error} (MSE) of the wideband noise signal in \Figref{fig:CRLB_vs_mean_error_wn}, 
it can be seen that the squared estimation error does not follow the CRLB as expected.
Also, this suggests that the source of the estimation error is mainly caused by model error.

\section{Conclusions}
We have proposed a method to estimate the DOA utilizing the directional sensitivity of microphones.
To show the potential, a sensor array with eight microphones has been used to estimate the DOA of four different types of signals. 
First, a model of the directivity pattern of each microphone was derived using a wideband noise signal emitted from different DOA angles.
Then, the received signal power at each microphone was compared with the model.
The result shows great potential, with a RMSE value around $1$\textdegree{} for all investigated signals.
Finally, we have analyzed the proposed approach using CRLB and found out that model error seems to be the main source of estimation error.

Future work includes experiments with other type of signals in order to further investigate the performance of the method. 
Also, the model error needs to be investigated and preferably included in the CRLB. 
Finally, it would be interesting to investigate how reverberations affect the algorithm, and whether they can be used to map the environment or localize the sound source.

\bibliographystyle{IEEEbib}
\bibliography{refs}

\begin{thebibliography}{10}

\bibitem{Knapp:DoaStudy}
C.~Knapp and G.~Carter,
\newblock ``The generalized correlation method for estimation of time delay,''
\newblock {\em IEEE Transactions on Acoustics, Speech, and Signal Processing},
  vol. 24, no. 4, pp. 320--327, 1976.

\bibitem{Krim:DoaStudy}
H.~Krim and M.~Viberg,
\newblock ``{Two decades of array signal processing research: the parametric
  approach},''
\newblock {\em IEEE Signal Processing Magazine}, vol. 13, no. 4, pp. 67--94,
  1996.

\bibitem{Stoica:DoaStudy}
P.~Stoica, P.~Babu, and J.~Li,
\newblock ``{SPICE: A Sparse Covariance-Based Estimation Method for Array
  Processing},''
\newblock {\em IEEE Transactions on Signal Processing}, vol. 59, no. 2, pp.
  629--638, 2011.

\bibitem{Rascon:RecentDoaStudy}
C.~Rascon and I.~Meza,
\newblock ``Localization of sound sources in robotics: A review,''
\newblock {\em Robotics and Autonomous Systems}, vol. 96, pp. 184--210, 2017.

\bibitem{Schmidt:MUSIC}
R.~Schmidt,
\newblock ``Multiple emitter location and signal parameter estimation,''
\newblock {\em IEEE Transactions on Antennas and Propagation}, vol. 34, no. 3,
  pp. 276--280, 1986.

\bibitem{Capon:MVDR)}
J.~Capon,
\newblock ``High-resolution frequency-wavenumber spectrum analysis,''
\newblock {\em Proceedings of the IEEE}, vol. 57, no. 8, pp. 1408--1418, 1969.

\bibitem{Buck:Directivity}
M.~Buck and M.~R{\"o}{\ss}ler,
\newblock ``First order differential microphone arrays for automotive
  applications,''
\newblock in {\em 7th International Workshop on Acoustic Echo and Noise
  Control}, Darmstadt, Germany, 2001.

\bibitem{Nakadai:Directivity}
K.~Nakadai, H.~Nakajima, K.~Yamada, Y.~Hasegawa, T.~Nakamura, and H.~Tsujino,
\newblock ``{Sound source tracking with directivity pattern estimation using a
  64 ch microphone array},''
\newblock in {\em 2005 IEEE/RSJ International Conference on Intelligent Robots
  and Systems}, Edmonton, AB, Canada, 2005, pp. 1690--1696.

\bibitem{Lofberg2004}
J.~L{\"{o}}fberg,
\newblock ``{YALMIP : A Toolbox for Modeling and Optimization in MATLAB},''
\newblock in {\em In Proceedings of the CACSD Conference}, Taipei, Taiwan,
  2004.

\bibitem{Ljung:SI}
L.~Ljung,
\newblock {\em System Identification: Theory for the User},
\newblock Prentice-Hall, Inc., 1986.

\bibitem{kay:estimation}
S.~M. Kay,
\newblock {\em {Fundamentals of Statistical Signal Processing: Estimation
  Theory}},
\newblock Prentice-Hall, Inc., 1993.

\bibitem{CBL99}
AKG,
\newblock ``{CBL99 Hemispherical boundary layer microphone},'' Accessed:
  2022-10-05.

\bibitem{UMC1820}
Behringer,
\newblock ``{U-PHORIA UMC1820},'' Accessed: 2022-10-05.

\bibitem{1029A}
Genelec,
\newblock ``{Genelec 1029A Data Sheet},'' 2003,
\newblock Accessed: 2022-10-05.

\end{thebibliography}

\end{document}